\documentclass[fleqn,10pt]{wlscirep}
\usepackage[utf8]{inputenc}
\usepackage[T1]{fontenc}
\usepackage{algorithm}
\usepackage{algpseudocode}
\usepackage{float}
\usepackage{physics}
\usepackage{amsthm}
\newtheorem{definition}{Definition}
\newtheorem{proposition}{Proposition}
\usepackage{multirow}

\title{Subspace-Confined QAOA with Generalized Dicke States for Multi-Channel Allocation in 5G CBRS Networks}

\author[1]{Gunsik Min}
\author[1]{Youngjin Seo}
\author[1,*]{Jun Heo}
\affil[1]{School of Electrical Engineering, Korea University, Seoul, 02841, Republic of Korea}
\affil[*]{junheo@korea.ac.kr}


\begin{abstract}
Efficient spectrum sharing in the Citizens Broadband Radio Service (CBRS) band is essential for maximizing 5G network capacity, particularly when high-traffic base stations require simultaneous access to multiple channels. Standard formulations of the Quantum Approximate Optimization Algorithm (QAOA) impose such multi-channel constraints using penalty terms, so most of the explored Hilbert space corresponds to invalid assignments. We propose a subspace-confined QAOA tailored to CBRS multi-channel allocation, in which each node-wise channel register is initialized in a Generalized Dicke state and evolved under an intra-register XY mixer. This ansatz confines the dynamics to a tensor product of Johnson graphs that exactly encode per-node Hamming-weight constraints. For an 8-node CBRS interference graph with 24 qubits, the effective search space is reduced from the full Hilbert space of size $2^{24}$ to 2{,}916 feasible configurations. Within this subspace, the algorithm converges rapidly to low-conflict assignments without large penalty coefficients. Simulations on instances with up to eight nodes show that the proposed ansatz achieves near-optimal conflict levels and consistently outperforms standard penalty-based QAOA and a greedy classical heuristic in terms of feasibility. Noise simulations with depolarizing channels further indicate that the constraint-preserving structure maintains a high feasibility ratio in NISQ-relevant error regimes.
\end{abstract}

\begin{document}

\flushbottom
\maketitle
\thispagestyle{empty}

\section*{Introduction}

Quantum computing offers algorithmic speedups for certain structured problems such as order-finding and search, and has motivated a wide range of quantum algorithms for optimization and simulation tasks\cite{BBHT, Deutsch85, DH96, DJ92, GR96, Hogg00,Shor94, Simon94, HHL09}. In the near term, however, gate fidelities and qubit counts limit us to noisy intermediate-scale quantum (NISQ) devices\cite{Preskill2018NISQ}, for which deep fault-tolerant algorithms are not yet practical. Variational quantum algorithms have therefore attracted considerable interest as a way to exploit shallow parameterized circuits combined with classical optimization\cite{Montanaro2016Overview,Cerezo2021VQA,Bharti2022NISQ}. Among them, the Quantum Approximate Optimization Algorithm (QAOA)\cite{QAOA} has emerged as a flexible framework for combinatorial optimization, with applications to MaxCut and related graph problems\cite{maxkcut,MAX18,QAOAmax,Eigen17,Eigen22,MA-QAOA}.

Standard QAOA implementations are typically built from a cost Hamiltonian that encodes the objective function and a simple mixer such as a transverse-field $X$ Hamiltonian. Hard constraints are then imposed through large penalty terms in the cost Hamiltonian. This approach is easy to formulate, but it forces the quantum state to explore the entire Hilbert space and relies on penalties to suppress invalid configurations. In highly constrained problems, only a tiny fraction of basis states are feasible, and the optimization landscape can become dominated by penalty contributions.

Recently, quantum algorithms have been proposed for problems in wireless communications, including MIMO detection and channel assignment\cite{weightedGrover22, QLSA22}. In this work, we consider CBRS channel allocation. The Federal Communications Commission (FCC) has reallocated 150~MHz of the 3.5~GHz band (3.55--3.7~GHz) to establish the CBRS band and promote spectrum sharing among military, satellite, and civilian users\cite{CBRSref1,CBRSref2,CBRSref3,CBRSref4}. CBRS users are divided into three tiers: incumbents, Priority Access License (PAL) holders, and General Authorized Access (GAA) users, with access coordinated by a Spectrum Access System (SAS). For PAL and GAA users, frequency allocation can be mapped to a graph coloring problem\cite{CBRS1,CBRS2}: CBSDs (Citizens Broadband Radio Service Devices) are represented as vertices, and interference constraints as edges. When a 150~MHz band is divided into 10~MHz channels, there are effectively 15 ``colors'' to assign. In realistic deployments, however, some CBSDs require more than 20~MHz, i.e.\ multiple channels simultaneously, so that each node must be assigned $k \ge 2$ distinct colors without conflict. The problem is then a graph multi-coloring instance with node-dependent cardinality constraints.

\begin{figure}
    \centering
    \begin{minipage}{.48\textwidth}
        \centering
        \includegraphics[width=\linewidth]{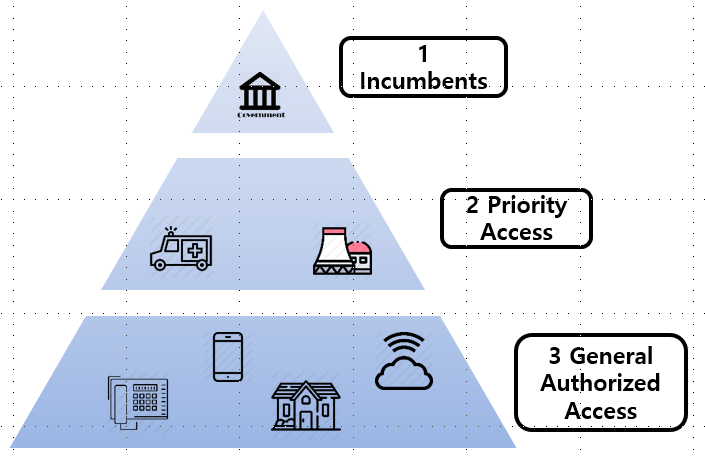}
        \caption*{(a)}
    \end{minipage}
    \hfill
    \begin{minipage}{.48\textwidth}
        \centering
        \includegraphics[width=\linewidth]{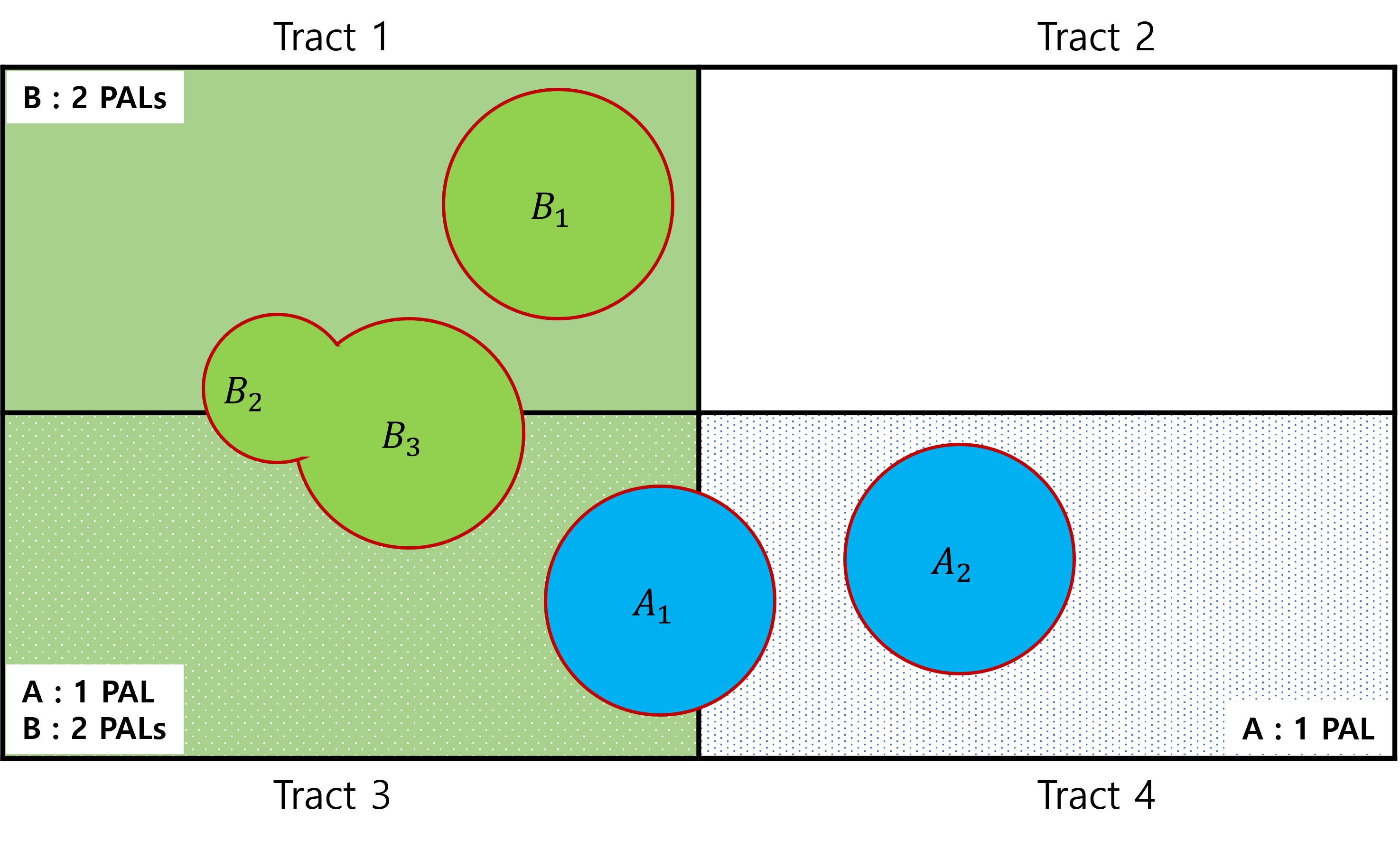}
        \caption*{(b)}
    \end{minipage}
    \caption{(a) Three-tier spectrum access in the 3.5~GHz CBRS band. (b) Example of PAL allocation over census tracts: PA~A holds one PAL covering tracts~1 and~3, while PA~B holds two PALs covering tracts~1 and~2. Each PA CBSD is constrained by a Priority Protection Area (PPA) that cannot extend beyond its service area.}
    \label{fig:CBRS}
\end{figure}

Figure~\ref{fig:CBRS} illustrates the three-tier access structure of the CBRS band and an example of PAL allocation across census tracts. Although the mapping to graph coloring is well known, incorporating realistic multi-channel demands and capacity constraints into quantum optimization algorithms is non-trivial. Most prior QAOA-based studies on coloring have focused on single-color assignments per node ($k=1$), sometimes using W-states and XY mixers to preserve a global Hamming weight, but without addressing heterogeneous multi-channel demands or quantifying search-space reduction in such settings.

\textbf{Related work.}
The connection between channel allocation, interference management, and graph coloring is well established in classical wireless communications as well as in recent quantum optimization studies\cite{CBRS1,CBRS2,maxkcut,MAX18}. Many QAOA-based approaches adopt penalty Hamiltonians combined with transverse-field mixers\cite{QAOAmax,Eigen17,Eigen22,EggerWarmStart2021}. XY-type mixers that preserve Hamming weight have been proposed to better handle constraints\cite{XYmixer}, but most analyses focus on single-coloring or on abstract benchmark graphs.

Constraint-preserving variants of QAOA based on Hamming-weight conserving XY mixers and Dicke (generalized W) states have been explored for graph coloring, portfolio optimization, and constrained summarization\cite{WangXYMixer,NiroulaXYQAOA,HeXYPortfolio,BrandhoferPortfolio,Hadfield2019QAOA,CookQAOAConstraints2019}. These works typically enforce a single global Hamming-weight constraint, for example selecting exactly $B$ assets from $n$ candidates or assigning one color per vertex, and often consider synthetic or small-scale problem instances.

In contrast, we consider a multi-register, node-dependent setting motivated by CBRS PAL allocation, in which each node $v_i$ carries a local register of $m$ channel qubits with its own demand $k_i \ge 2$. The feasible space is the direct product of heterogeneous Dicke subspaces. We design a subspace-confined ansatz that enforces these node-wise multi-coloring constraints directly in the mixer, and we characterize the resulting feasible subspace in terms of tensor products of Johnson graphs. At the level of dynamics, the XY mixer realizes a quantum walk on this tensor-product Johnson graph, while the cost Hamiltonian encodes interference conflicts.

\textbf{Our contribution.}
We formulate PAL channel assignment in CBRS as a graph multi-coloring problem with node-dependent cardinality constraints that capture realistic multi-channel demands ($k_i \ge 2$). On top of this formulation, we propose a subspace-confined QAOA ansatz that combines node-wise Generalized Dicke state initialization with an intra-register XY-Hamiltonian mixer that preserves the Hamming weight of each node. This construction induces a direct-product structure of heterogeneous Dicke subspaces, which we describe using Johnson graphs, and restricts the QAOA dynamics to configurations that satisfy all per-node channel-demand constraints.

We quantitatively analyze the resulting search-space reduction and demonstrate, on CBRS interference graphs of up to 24 qubits, that the proposed ansatz converges to multi-coloring solutions that are competitive with exact integer-linear programming (ILP) and strictly better than a simple greedy heuristic, while maintaining a unit feasibility ratio even when standard penalty-based QAOA almost never produces valid samples. We also investigate robustness under depolarizing noise and observe that the constraint-preserving dynamics maintain a high feasibility ratio in NISQ-relevant noise regimes. Finally, we discuss a dual-constraint extension of the ansatz that simultaneously enforces node-wise channel demands and per-channel capacity constraints at the mixer level, illustrating how additional structure can be incorporated into the QAOA framework for more realistic CBRS deployment scenarios.

\section*{Results}

\subsection*{Subspace-confined QAOA: performance and noise robustness}

We first evaluate the proposed subspace-confined QAOA on an 8-node (24-qubit)
CBRS interference graph. The graph $G=(V,E)$ is constructed as an 8-node ring
with two additional cross edges to emulate realistic interference patterns
among eight CBSDs. We consider three available channels ($m=3$) and
heterogeneous node demands
\[
    k = [2, 1, 2, 1, 1, 2, 1, 1],
\]
so that the total number of qubits is $N_{\mathrm{qubits}} = n m = 24$.

\subsubsection*{Subspace confinement via node-wise Johnson graphs}

We begin by formalizing the structure of the feasible state space explored by our ansatz.

\begin{definition}[Node-wise Johnson-graph structure of the feasible subspace]
For each node $v_i$, we use a one-hot register of $m$ qubits subject to the
hard cardinality constraint
\begin{equation}
    \sum_{c=1}^{m} x_{i,c} = k_i.
\end{equation}
The computational basis states satisfying this constraint are in one-to-one
correspondence with the $k_i$-subsets of $\{1,\dots,m\}$, which form the
vertex set of the Johnson graph $J(m,k_i)$, whose edges connect pairs of
$k_i$-subsets that differ by exactly one element.
\end{definition}

In our subspace-confined ansatz, the feasible state space factorizes as the
tensor product
\begin{equation}
    \mathcal{H}_{\mathrm{feas}} \cong \bigotimes_{i=1}^{n} J(m,k_i),
\end{equation}
and its dimension is
\begin{equation}
    |\mathcal{F}| = \prod_{i=1}^{n} \binom{m}{k_i},
\end{equation}
where $\mathcal{F}$ denotes the set of bitstrings that satisfy all node-wise
Hamming-weight constraints. By contrast, the full Hilbert space of the
$N_{\mathrm{qubits}} = n m$ qubits has size $2^{nm}$.

\begin{proposition}[Search-space reduction from Johnson-graph confinement]
For the 8-node CBRS instance with $m=3$ channels and
$k = [2, 1, 2, 1, 1, 2, 1, 1]$, the restriction to the tensor-product
Johnson-graph subspace reduces the effective search space from
\begin{equation}
    2^{24} \approx 1.68\times 10^{7}
\end{equation}
to
\begin{equation}
    |\mathcal{F}| = \prod_{i=1}^{8} \binom{3}{k_i} = 2{,}916,
\end{equation}
corresponding to a search-space reduction factor of approximately
$5.8\times 10^{3}$.
\end{proposition}

In other words, the proposed Dicke+XY ansatz explores only the tensor-product
Johnson-graph subspace $\mathcal{H}_{\mathrm{feas}}$, while standard
penalty-based QAOA with an $X$-mixer evolves in the full Hilbert space,
where only a fraction $\approx 1.7\times 10^{-4}$ of states are feasible.

Table~\ref{tab:reduction} summarizes this search-space reduction for the
24-qubit instance.

\begin{table}[t]
\centering
\caption{Comparison of search-space dimensions for the 8-node CBRS instance
with $m=3$ channels (24 qubits). The proposed ansatz explores only the
tensor-product Johnson-graph subspace induced by the node-wise Hamming-weight
constraints.}
\label{tab:reduction}
\begin{tabular}{lccc}
\hline
\textbf{Method} & \textbf{Total states} & \textbf{Feasible states} & \textbf{Reduction factor} \\ \hline
Standard QAOA (X-mixer) & $2^{24} \approx 1.68\times 10^{7}$ & $2{,}916$ (fraction $\approx 1.7\times10^{-4}$) & 1 \\ 
Proposed (subspace-confined) & $2{,}916$ & $2{,}916$ (100\% of explored states) & $\approx 5.8\times 10^{3}$ \\ \hline
\end{tabular}
\end{table}

\subsubsection*{Performance comparison on an 8-node CBRS instance}

\begin{figure}[htp!]
    \centering
    \includegraphics[width=0.48\linewidth]{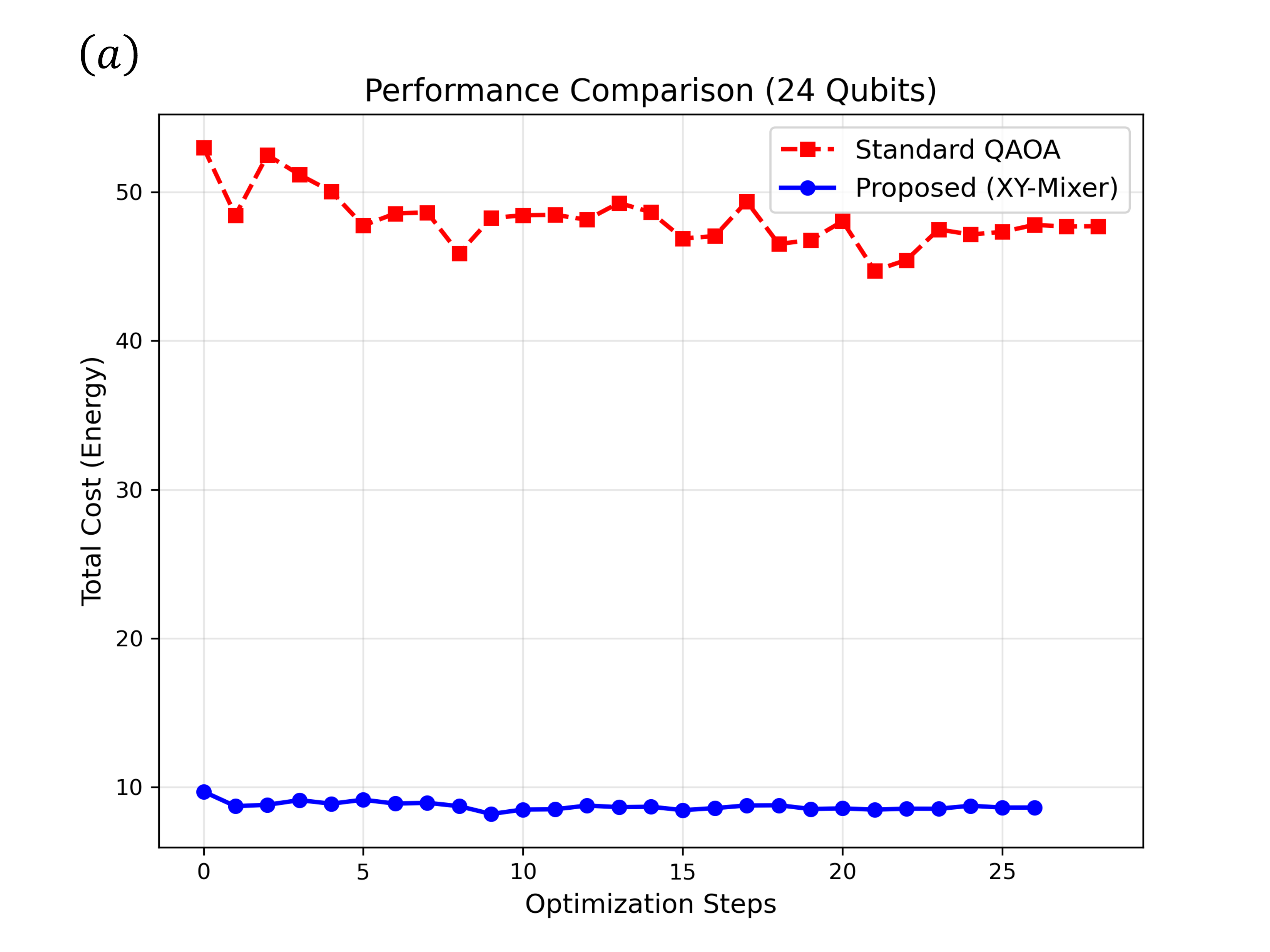}
    \includegraphics[width=0.42\linewidth]{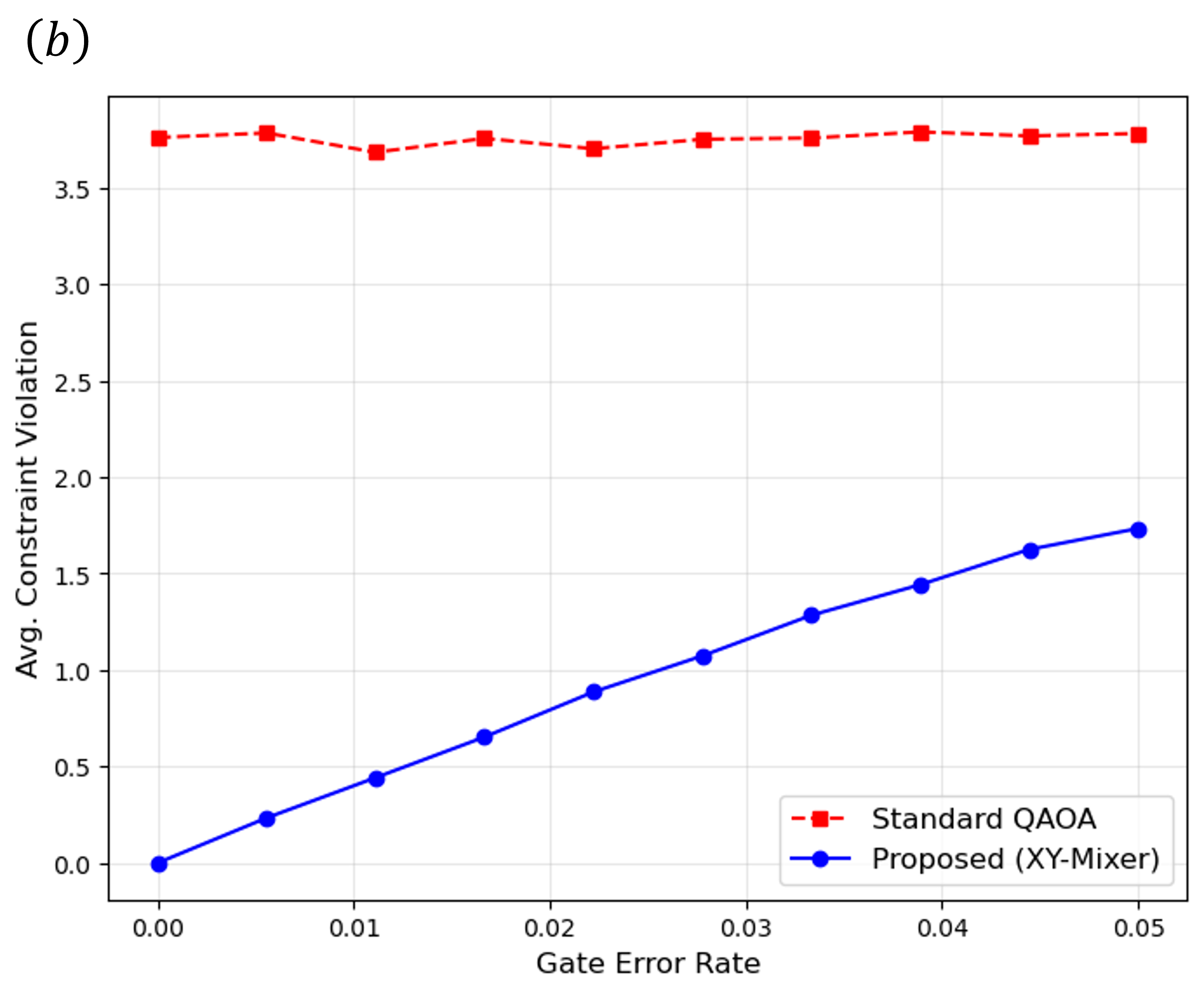}
    \caption{(a) Optimization trajectories for the 8-node CBRS network
    (24 qubits) comparing standard penalty-based QAOA and the proposed
    subspace-confined QAOA. (b) Average deviation from the Hamming-weight
    constraints as a function of depolarizing error rate for a 5-node
    (15-qubit) instance.}
    \label{fig:performance}
\end{figure}

Figure~\ref{fig:performance}(a) shows representative optimization trajectories
for the 8-node (24-qubit) instance at QAOA depth $p=1$. The standard QAOA
baseline uses an $X$-mixer and a penalty Hamiltonian with coefficient
$\lambda$ to enforce the demand constraints, whereas the proposed ansatz uses
node-wise Generalized Dicke state initialization and an intra-register
XY-mixer that preserves the Hamming weight of each node.

Consistent with the search-space analysis above, the standard QAOA remains
trapped in high-cost regions: its average cost (conflicts plus penalties)
saturates at a large value and it rarely samples feasible bitstrings. In
contrast, the subspace-confined QAOA starts directly inside
$\mathcal{H}_{\mathrm{feas}}$ and the XY-mixer guarantees that all sampled
states satisfy the node-wise cardinality constraints, so the optimizer can
focus solely on reducing interference conflicts.

On the 8-node CBRS instance, the classical ILP baseline finds an optimal
solution with two interference conflicts, while a simple greedy multi-coloring
heuristic returns a solution with four conflicts. The standard QAOA attains a
best feasible conflict of four, but the feasibility ratio---the fraction of
measurement shots that satisfy all node-wise Hamming-weight constraints—is
only $10^{-3}$. In contrast, the proposed subspace-confined QAOA achieves a
best feasible conflict of three with a feasibility ratio of $1.0$, i.e.\ all
sampled bitstrings satisfy the cardinality constraints. Thus, on this
non-trivial 24-qubit instance, the proposed ansatz both concentrates
probability mass within the feasible subspace and finds solutions that are
closer to the classical optimum than those obtained by the greedy heuristic
and the penalty-based QAOA.

\subsubsection*{Robustness under depolarizing noise}

We next investigate the robustness of the subspace-confined ansatz under
depolarizing noise on a smaller 5-node (15-qubit) CBRS instance, where noisy
simulations are more tractable. A standard depolarizing channel
with gate error rate $p_{\mathrm{err}}$ is applied to all single- and two-qubit
gates in the QAOA circuits.

As a performance metric, we consider the average deviation of the realized
Hamming weights from the node-wise demands,
\begin{equation}
    \text{Deviation} =
    \biggl\langle
        \sum_{v_i \in V}
        \Bigl|
            \sum_{c=1}^{m} x_{i,c} - k_i
        \Bigr|
    \biggr\rangle,
\end{equation}
where the average is taken over many measurement shots. A deviation of $1.0$
indicates that, on average, the total channel allocation across the network
differs by one channel from a valid configuration.

Figure~\ref{fig:performance}(b) plots this deviation as a function of
$p_{\mathrm{err}}$ for both the standard and proposed QAOA ansätze. The
standard penalty-based QAOA exhibits a consistently high deviation
($\approx 1.5$--$2.0$), indicating that it predominantly samples invalid
solutions regardless of the noise level. In contrast, the subspace-confined
ansatz starts with near-zero deviation at $p_{\mathrm{err}}=0$ and increases
approximately linearly with $p_{\mathrm{err}}$, remaining below $1.0$ even at
the largest error rates considered. These results suggest that the structural
protection offered by the XY-mixer and Dicke-state initialization provides
inherent robustness against depolarizing noise in NISQ-relevant regimes.

\begin{figure}[t]
    \centering
    \includegraphics[width=0.5\linewidth]{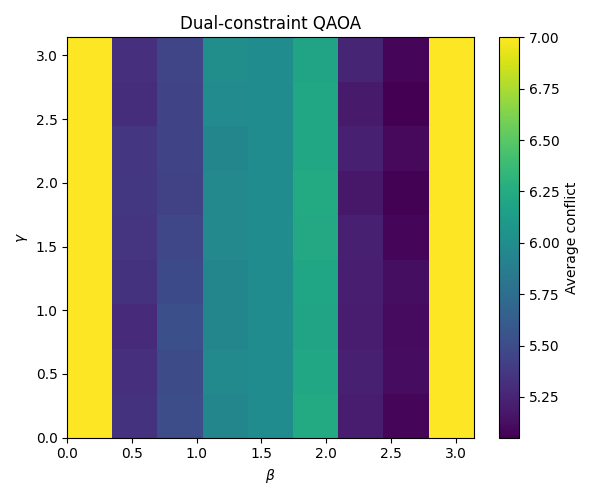}
    \caption{Average conflict landscape for the dual-constraint
    QAOA ansatz on an 8-node, 3-channel CBRS interference graph
    (24 qubits). The color scale indicates the average number of
    interference conflicts as a function of the problem and mixer
    angles $(\gamma,\beta)$ at depth $p=1$. For all parameter pairs
    considered, both the node-wise and channel-wise cardinality
    constraints remain satisfied with probability close to one.}
    \label{fig:dual_heatmap}
\end{figure}

\subsection*{Dual-constraint experiment on a CBRS instance}

To illustrate how the dual-constraint ansatz behaves in a realistic
setting, we again consider an 8-node CBRS interference graph with
three channels ($n=8$, $m=3$), requiring $N_{\mathrm{qubits}} = 24$.
The node-wise channel demands are
\[
    k = [2, 1, 2, 1, 1, 2, 1, 1],
\]
and we impose fixed per-channel capacities
\[
    L = [4, 4, 3],
\]
so that $\sum_i k_i = \sum_c L_c = 11$. The interference edges follow
an 8-node ring with two additional cross links, matching the
topology used in the main subspace-confined experiments.

Using the deterministic greedy procedure described in the Methods
section, we first construct a canonical dual-feasible matrix
$X^{(0)}$ satisfying both the $k_i$ and $L_c$ constraints, and prepare
the corresponding basis state $\ket{X^{(0)}}$ on the 24-qubit
register. We then apply a single QAOA layer ($p=1$) with the
conflict-counting problem Hamiltonian $H_P$ and the plaquette-based
dual mixer $\exp(-i \beta H_{\mathrm{mix}}^{\mathrm{dual}})$, and
perform a grid scan over $(\gamma,\beta) \in [0,\pi]\times[0,\pi]$ on
a $9 \times 9$ mesh. For each parameter pair we run the circuit on a
noiseless \texttt{qasm\_simulator} backend with 2{,}048 shots and
estimate the average number of interference conflicts as well as the
fractions of samples that satisfy the node-wise and channel-wise
cardinality constraints.

Figure~\ref{fig:dual_heatmap} shows a representative heatmap of the
average conflict as a function of $(\gamma,\beta)$: darker regions
correspond to lower-conflict allocations. Across the scanned grid, the
node-feasibility and channel-feasibility ratios remain essentially
equal to one, confirming that the dual mixer preserves both sets of
cardinality constraints at the circuit level. Within this strictly
feasible subspace, we observe parameter regions where the average
conflict is comparable to the ILP optimum and strictly better than a
simple greedy multi-coloring heuristic, even at depth $p=1$.

\subsection*{Scalability and comparison with classical baselines}

To assess scalability and compare explicitly with classical methods, we
consider CBRS interference graphs with $N \in \{6,7,8\}$ nodes, three channels
($m=3$), and heterogeneous demands
$k = [2, 1, 2, 1, 1, 2, 1, 1]$ truncated to the first $N$ nodes. For each
instance, we compute an exact classical optimum using a 0--1 ILP over the
feasible subspace, a simple greedy multi-coloring heuristic, and depth-$p=1$
QAOA with either the standard $X$-mixer and penalty Hamiltonian or the
proposed Dicke+XY mixer ansatz.

\begin{table}[htp!]
\centering
\caption{Optimality gap (best conflict minus ILP optimum) and feasibility
ratio for classical baselines and QAOA ansätze on CBRS interference graphs
with $N \in \{6,7,8\}$ nodes. A gap of~0 indicates that the method matches the
ILP optimum; ``n/a'' indicates that no feasible sample was observed.}
\label{tab:gap_feas}
\begin{tabular}{ccccccc}
\hline
\textbf{\#nodes} &
\multicolumn{2}{c}{\textbf{Greedy}} &
\multicolumn{2}{c}{\textbf{Standard QAOA}} &
\multicolumn{2}{c}{\textbf{Proposed QAOA}} \\
\cline{2-7}
& gap & feas. & gap & feas. & gap & feas. \\
\hline
6 & $4-3 = 1$ & 1.000 & $4-3 = 1$ & 0.005 & $3-3 = 0$ & 1.000 \\
7 & $3-2 = 1$ & 1.000 & n/a       & 0.000 & $2-2 = 0$ & 1.000 \\
8 & $4-2 = 2$ & 1.000 & $4-2 = 2$ & 0.001 & $3-2 = 1$ & 1.000 \\
\hline
\end{tabular}
\end{table}

\begin{table}[t]
\centering
\caption{Comparison of classical baselines and QAOA ansätze on CBRS
interference graphs with $m=3$ channels and heterogeneous node demands
($k = [2,1,2,1,1,2,1,1]$ truncated to $N$ nodes). ``Best conflict'' denotes
the minimum number of interference conflicts observed; the feasibility ratio
for quantum methods is the fraction of measurement shots that satisfy all
node-wise Hamming-weight constraints. For $N=7$, the standard QAOA did not
produce any feasible samples within the optimization budget.}
\label{tab:classical_qaoa}
\begin{tabular}{ccccccc}
\hline
\textbf{\#nodes} &
\textbf{ILP best} &
\textbf{Greedy best} &
\textbf{Std QAOA best} &
\textbf{Std feas.} &
\textbf{Prop QAOA best} &
\textbf{Prop feas.} \\
\hline
6 & 3 & 4 & 4 & 0.005 & 3 & 1.000 \\
7 & 2 & 3 & -- & 0.000 & 2 & 1.000 \\
8 & 2 & 4 & 4 & 0.001 & 3 & 1.000 \\
\hline
\end{tabular}
\end{table}

The greedy heuristic shows a non-zero optimality gap on all tested sizes, whereas the proposed QAOA either matches the ILP optimum ($N=6,7$) or stays within one conflict of it ($N=8$), with feasibility ratio essentially equal to one. The penalty-based QAOA, in contrast, almost never samples feasible solutions in this regime, and for $N=7$ it does not yield any feasible sample at all within the given optimization budget. For the CBRS multi-channel allocation instances considered here, the subspace-confined Dicke+XY ansatz therefore provides near-optimal multi-colorings that are competitive with exact ILP solutions and strictly better than a simple greedy heuristic, while preserving the constraint structure with unit feasibility ratio as the network size grows.

\section*{Discussion}

The numerical results on CBRS-inspired interference graphs show that
explicit subspace confinement can substantially affect QAOA performance
for multi-channel allocation problems. For the 8-node (24-qubit)
instance, the Dicke+XY ansatz restricts the dynamics to a tensor
product of node-wise Johnson graphs and reduces the size of the
explored state space by a factor of about $5.8\times 10^{3}$ compared
with the full Hilbert space. Within this subspace all sampled
bitstrings automatically satisfy the per-node demand constraints, so
the optimization focuses on reducing interference conflicts. In
contrast, the standard penalty-based QAOA with an $X$-mixer explores a
space in which only a very small fraction of states are feasible and
the penalty terms dominate the landscape. In our tests, this leads to
optimization trajectories that only rarely visit valid configurations,
even at shallow depth.

The comparison with classical baselines clarifies the practical impact
of this behavior. On small CBRS instances with
$N \in \{6,7,8\}$ nodes and three channels, an exact ILP solver finds
the global optimum over the feasible subspace, while a simple greedy
multi-coloring heuristic gives solutions with one or two additional
conflicts. The proposed subspace-confined QAOA either matches the ILP
optimum (for six and seven nodes) or remains within one conflict of it
(for eight nodes), and in all cases produces feasible bitstrings with
probability close to one. The standard penalty-based QAOA behaves very
differently: the feasibility ratio is essentially zero for seven nodes and
of order $10^{-3}$ for six and eight nodes, and even when feasible
bitstrings appear their probability is extremely small. For the CBRS
setting considered here, the main advantage of the subspace-confined
ansatz is therefore not a dramatic improvement in approximation ratio
over classical methods, but the ability to reliably generate valid
channel assignments with competitive conflict levels without tuning
large penalty coefficients.

The noise-robustness study on a 5-node (15-qubit) instance supports
this interpretation. When depolarizing noise is applied to all gates,
the standard penalty-based QAOA shows a large and almost
noise-independent deviation from the target Hamming weights, which is
consistent with the fact that it already concentrates weight in the
invalid part of the Hilbert space. The subspace-confined ansatz
starts with very small deviation in the ideal case and then increases
roughly linearly with the error rate, remaining below one effective
channel error over the range considered. This suggests that a
constraint-preserving mixer can provide a degree of structural
protection for shallow circuits on noisy devices, although the extent
of this benefit will depend on the noise model and hardware.

Beyond the node-wise Dicke+XY construction, we also explored a
dual-constraint variant of the ansatz that enforces both per-node
channel demands and per-channel capacities through a plaquette-based
mixer. In this setting, feasible configurations must satisfy
$\sum_c x_{i,c} = k_i$ for all nodes and
$\sum_i x_{i,c} = L_c$ for all channels. Starting from a deterministically
constructed basis state $\ket{X^{(0)}}$ that satisfies both
constraints, the dual mixer is built from local four-qubit unitaries
that coherently exchange channel allocations between pairs of nodes
and pairs of channels while preserving all row and column sums in the
allocation matrix. On an 8-node, 3-channel instance (24 qubits) with
fixed per-channel capacities, a shallow dual-constraint QAOA circuit
($p=1$) explores only the intersection of the node- and
channel-cardinality subspaces. In a two-parameter sweep over the
problem and mixer angles, the node- and channel-feasibility
ratios remain essentially equal to one for all parameter pairs,
confirming that the dual mixer preserves both sets of constraints at
the circuit level. Within this feasible region, we observe parameter
choices for which the average conflict is comparable to the ILP
optimum and strictly better than a simple greedy multi-coloring
heuristic, even without penalty terms in the problem Hamiltonian. This
example illustrates how additional domain structure, such as
per-channel capacity limits, can be incorporated directly at the mixer
level in a way that remains compatible with the overall QAOA framework.

From the viewpoint of wireless communications, the ability to handle
heterogeneous multi-channel demands in CBRS is important for modelling
realistic PAL deployments. Many existing quantum formulations for
channel allocation implicitly assume a single channel per node or a
single global cardinality constraint, which does not reflect
high-traffic CBSDs that require multiple channels. In our framework,
node-wise Generalized Dicke states and intra-register XY mixers enforce
per-node multi-coloring constraints by construction, and the feasible
configurations can be described as a tensor product of Johnson graphs.
The dual-constraint extension further captures fixed per-channel
capacities and brings the quantum model closer to practical
spectrum-allocation scenarios in which both node-side demand and
channel-side load must be controlled. This separation between local
cardinality constraints and global interference structure is simple,
yet flexible enough to be adapted to other resource allocation tasks
such as multi-carrier scheduling, beam assignment, or network slicing,
where similar combinatorial structures appear.

There are, however, several limitations of the present study. The
exact preparation of Generalized Dicke states increases circuit depth
compared to product-state initializations. For the problem sizes
considered here the benefit of staying in the feasible subspace
appears to outweigh this overhead, but for larger or noisier devices
the trade-off may become less favorable. Approximate Dicke-state
preparation and hardware-efficient decompositions of the XY and
plaquette mixers are natural directions for future work. In addition,
all numerical experiments in this study are limited to at most 24
qubits and use simplified interference graphs. Real CBRS deployments
involve larger topologies, more channels, and additional constraints
such as coexistence with GAA users, realistic propagation models, and
time-varying traffic. Extending the present approach to such settings
will likely require a combination of problem decomposition, classical
preprocessing, and heuristic parameter initialization.

Finally, the classical baselines used here are intentionally simple: an
exact ILP over the feasible subspace and a straightforward greedy
heuristic. These are sufficient to place the quantum results in
context, but do not exhaust the range of classical methods for graph
multi-coloring. A more systematic comparison with advanced
metaheuristics or local-search algorithms would be useful for
assessing the potential of subspace-confined and dual-constraint QAOA
on larger instances. Small-scale hardware experiments with the
Dicke+XY and dual-constraint ansätze would also help to clarify how
calibration errors, device-specific noise, and shot limitations affect
the behavior observed in simulation. Overall, our results indicate
that tailoring variational ansätze to the constraint structure of
multi-channel spectrum allocation---including both per-node demands and
per-channel capacities---can make QAOA more relevant for realistic
CBRS-type scenarios, even on near-term devices.

\section*{Methods}

\subsection*{Problem formulation as a graph multi-coloring instance}

We formulate the CBRS channel allocation task as a Graph Multi-Coloring Problem (GMCP)\cite{GMCPRef}. Each certified CBSD is represented as a vertex in an interference graph $G=(V,E)$ with $n$ nodes, and each available frequency channel is treated as a color from the set $K = \{c_1,\dots,c_m\}$. An interference link between two CBSDs appears as an edge $(v_i,v_j) \in E$, indicating that these two nodes cannot share the same channel.

We employ a one-hot encoding in which node $v_i$ is described by $m$ binary variables $x_{i,c}$, where $x_{i,c}=1$ if channel $c$ is assigned to $v_i$ and $x_{i,c}=0$ otherwise. Unlike the standard single-coloring case with $\sum_c x_{i,c}=1$, we allow each node $v_i$ to request a node-dependent number $k_i$ of channels. Valid channel allocations must satisfy two types of constraints. First, the demand constraint requires that each node receives exactly $k_i$ channels,
\begin{equation}
    \sum_{c=1}^{m} x_{i,c} = k_i \quad \forall v_i \in V,
\end{equation}
which implies that the quantum state encoding node $v_i$ must have Hamming weight $k_i$. Second, the interference constraint enforces that adjacent nodes cannot use the same channel simultaneously,
\begin{equation}
    x_{i,c} \cdot x_{j,c} = 0 \quad \forall (v_i, v_j) \in E,\ \forall c \in K.
\end{equation}

To map this problem to a quantum system, we allocate $N_{\mathrm{qubits}} = n m$ qubits, one for each binary decision variable $x_{i,c}$. The presence of multi-channel demands ($k_i \ge 2$) significantly reduces the relative density of valid solutions in the total Hilbert space. As $n$ and the $k_i$ increase, the fraction of basis states that satisfy all constraints decreases rapidly, which motivates the subspace-confined QAOA approach described below.

\subsection*{Standard QAOA baseline}

The Quantum Approximate Optimization Algorithm (QAOA)\cite{QAOA} is a hybrid quantum--classical algorithm designed to approximate the minimum of a classical cost function $C(z)$ defined on bitstrings $z$. The cost is encoded in a problem Hamiltonian $H_P$ whose eigenvalues reproduce $C(z)$, and the algorithm searches over a family of parameterized quantum states.

A depth-$p$ QAOA circuit prepares a trial state
\begin{equation}
    |\psi(\boldsymbol{\gamma}, \boldsymbol{\beta})\rangle 
    = \prod_{l=1}^p e^{-i\beta_l H_M} e^{-i\gamma_l H_P} \ket{+}^{\otimes N_{\mathrm{qubits}}},
\end{equation}
where $H_P$ encodes the cost function, $H_M$ is the mixing Hamiltonian, and $\ket{+}^{\otimes N_{\mathrm{qubits}}}$ is the uniform superposition over all computational basis states. In the standard formulation, the mixer is taken to be the transverse-field Hamiltonian $H_M = \sum_{j} \sigma_x^j$.

Applied directly to CBRS multi-coloring, this baseline faces considerable difficulty. The X-mixer allows the state to explore the full Hilbert space, including configurations that violate the per-node demand constraints. To discourage such violations, one typically augments the problem Hamiltonian with quadratic penalties,
\begin{equation}
    H_P = H_{\text{cost}} 
    + \lambda \sum_{v_i \in V} 
    \left( \sum_{c=1}^{m} \hat{n}_{i,c} - k_i \right)^2,
    \label{eq:penalty}
\end{equation}
where $\lambda$ is a large penalty coefficient and 
$\hat{n}_{i,c} = \tfrac{1}{2}(I - \sigma_{z}^{(i,c)})$ 
is the number operator on qubit $(i,c)$. These penalties make the landscape more rugged and can introduce many local minima, and in practice the performance can be very sensitive to the choice of $\lambda$\cite{Eigen22,MA-QAOA}.

For an $n$-node, $m$-channel CBRS instance, the number of feasible configurations that satisfy all Hamming-weight constraints is
\begin{equation}
    \left|\mathcal{F}\right|
    = \prod_{i=1}^{n} \binom{m}{k_i}.
\end{equation}
By comparison, the full Hilbert space has size $2^{N_{\mathrm{qubits}}}$. In our 8-node, 3-channel example with $N_{\mathrm{qubits}} = 24$ and 
$k = [2,1,2,1,1,2,1,1]$, there are $2{,}916$ feasible states out of $2^{24} \approx 1.68\times 10^{7}$ possible bitstrings, corresponding to a fraction of about
\[
    \frac{\left|\mathcal{F}\right|}{2^{24}} \approx 1.7\times10^{-4}.
\]
Only a tiny portion of the state space therefore represents valid channel allocations. The standard X-mixer spends most of its evolution exploring invalid assignments, and the penalties in Eq.~\eqref{eq:penalty} must be chosen large enough to offset this effect. In such regimes, barren-plateau phenomena, where gradients vanish exponentially with system size, can be exacerbated\cite{BarrenPlateau}. Empirically, we observe that for the 24-qubit instance considered here, standard QAOA with a transverse-field mixer and the penalty Hamiltonian~\eqref{eq:penalty} fails to reach low-energy feasible solutions within a reasonable optimization budget.

In all our numerical experiments, we focus on shallow circuits with depth $p=1$ for both the standard and subspace-confined ansätze. This depth is sufficient to highlight the qualitative performance differences on the CBRS instances studied. We use the same classical optimizer (COBYLA), the same random initial parameters, and set $\lambda = 5.0$ in Eq.~\eqref{eq:penalty}, verifying that moderate changes of $\lambda$ do not qualitatively change the observed behavior.

\subsection*{Subspace-confined QAOA ansatz}

\subsubsection*{Subspace-preserving evolution via XY-Hamiltonian}

To address the strict cardinality constraints, we replace the transverse-field mixer with an XY-Hamiltonian that preserves the total Hamming weight on each node. For a system with $N_{\mathrm{qubits}}$ qubits, the mixer is constructed as
\begin{equation}
    H_{XY} = \frac{1}{2} \sum_{(i,j) \in E_{\text{mix}}} 
    \left(\sigma_i^x \sigma_j^x + \sigma_i^y \sigma_j^y\right),
\end{equation}
where $\sigma^x$ and $\sigma^y$ are Pauli operators and $E_{\text{mix}}$ is a set of qubit pairs within the same node (intra-node connectivity) that are allowed to exchange excitations.

Writing this Hamiltonian in terms of lowering and raising operators,
\begin{equation}
    H_{XY} = \sum_{(i,j) \in E_{\text{mix}}} 
    \left(\sigma_i^+ \sigma_j^- + \sigma_i^- \sigma_j^+\right)
    = \sum_{(i,j) \in E_{\text{mix}}} 
    \left(\ketbra{1_i 0_j}{0_i 1_j} + \ketbra{0_i 1_j}{1_i 0_j}\right),
\end{equation}
we see that it acts as a partial SWAP gate: it exchanges excitations between two qubits within the same node while conserving the total Hamming weight. The mixer commutes with the total number operator on each node, so if the initial state lies in a fixed-weight subspace, the evolution $e^{-i\beta H_{XY}}$ remains in that subspace.

For the $m$-qubit register of a single node $v_i$ with Hamming weight $k_i$, the computational basis states with weight $k_i$ form the vertex set of the Johnson graph $J(m,k_i)$, and $H_{XY}$ induces transitions between pairs of basis states that differ by exchanging a single channel. On the full system, the mixer therefore implements a quantum walk on the tensor-product graph $\bigotimes_i J(m,k_i)$, which encodes all node-wise channel-demand constraints.

\begin{figure}[htbp]
    \centering
    \includegraphics[width=0.9\linewidth]{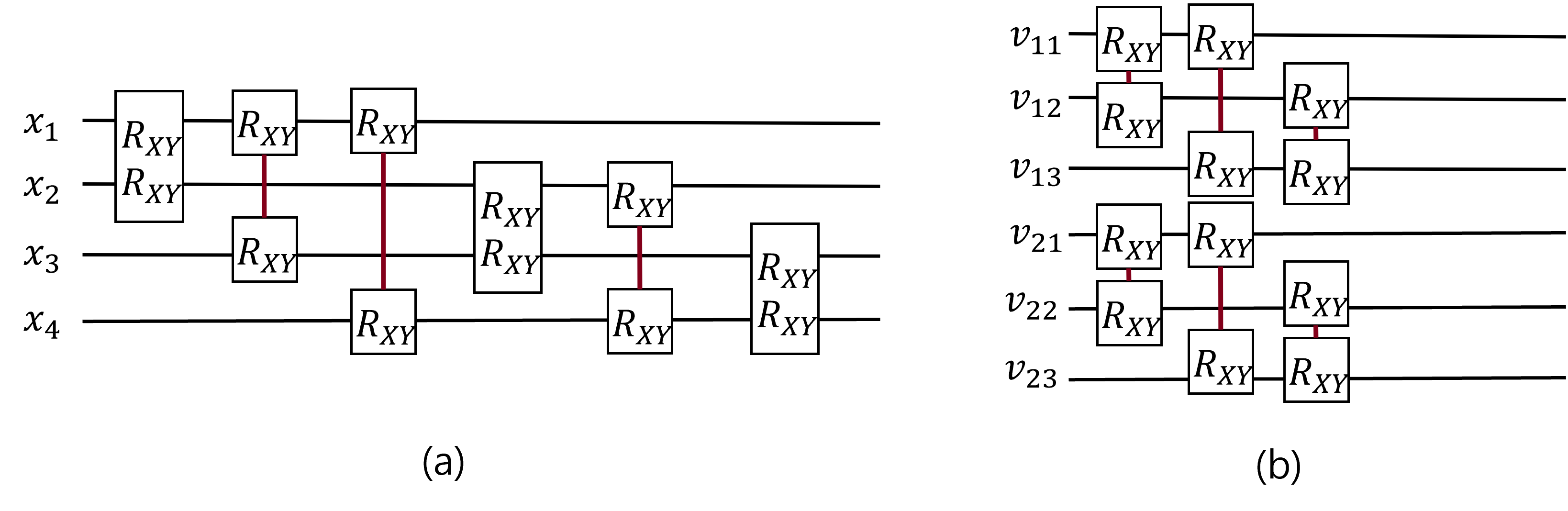}
    \caption{(a) Unitary evolution of the XY-Hamiltonian for a single node with four available colors. (b) Unitary evolution for a two-node, three-colorable problem. The mixer acts on intra-node qubit pairs and preserves the Hamming weight for each node, implementing a quantum walk on the corresponding Johnson graphs.}
    \label{fig:XY}
\end{figure}

Figure~\ref{fig:XY} illustrates the unitary evolution under the XY-Hamiltonian for single- and two-node examples. In each case, the mixer acts only on intra-node qubit pairs and explores different channel assignments without changing the number of assigned channels per node.

\subsubsection*{Deterministic initialization with Generalized Dicke states}

To fully exploit the structure-preserving property of the XY mixer, the initial state should be supported only on valid channel assignments that satisfy the demand $k_i$ for each node $v_i$. We therefore use Generalized Dicke states, denoted $|D_k^m\rangle$, which are equal superpositions over all computational basis states of $m$ qubits with Hamming weight $k$,
\begin{equation}
    |D_k^m\rangle = \frac{1}{\sqrt{\binom{m}{k}}} \sum_{x \in \{0,1\}^m,\ \mathrm{wt}(x)=k} |x\rangle.
\end{equation}

The global initial state is constructed as a tensor product of node-wise Dicke states,
\begin{equation}
    |\psi\rangle_{\mathrm{init}} = \bigotimes_{i=1}^{n} |D_{k_i}^m\rangle_{v_i},
\end{equation}
where $|D_{k_i}^m\rangle_{v_i}$ is the state for node $v_i$ with $m$ available channels and demand $k_i$. By construction, $|\psi\rangle_{\mathrm{init}}$ has support only on configurations that satisfy all node-wise Hamming-weight constraints. Under the XY mixer, the evolution remains in the tensor-product Johnson-graph subspace $\mathcal{H}_{\mathrm{feas}} \cong \bigotimes_{i} J(m,k_i)$ for all QAOA layers.

\begin{figure}[htbp]
    \centering
    \includegraphics[width=0.8\linewidth]{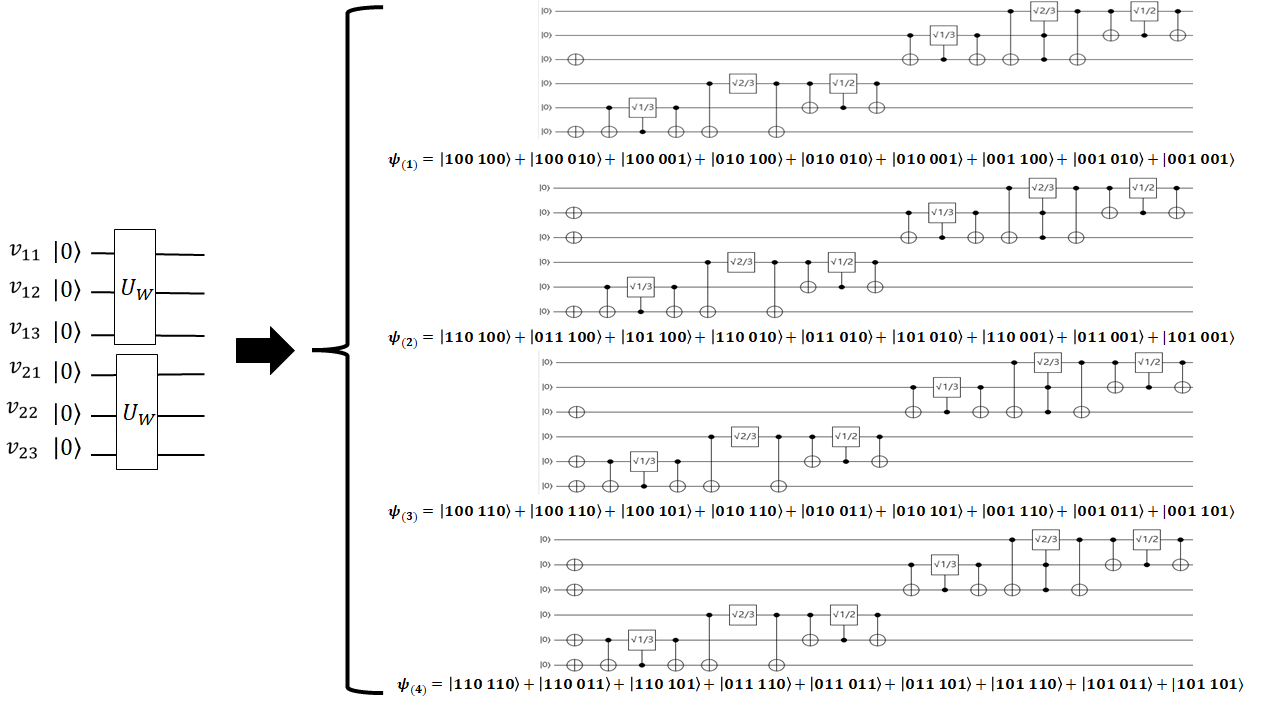}
    \caption{Quantum circuit for preparing the initial state in a two-node, three-channel system. The gate labeled $U_W$ generates the Generalized Dicke state $|D_k^m\rangle$. X gates at the input can be used to adjust the target Hamming weight $k_i$.}
    \label{fig:w-state}
\end{figure}

\begin{figure}[htbp]
    \centering
    \includegraphics[width=0.8\linewidth]{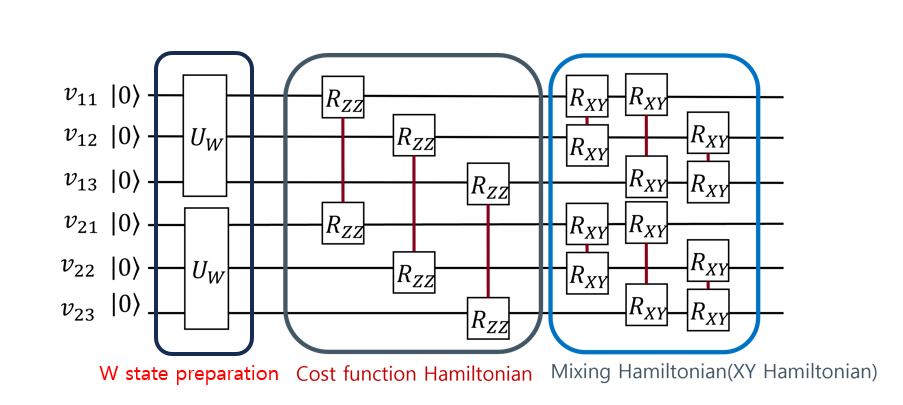}
    \caption{Overview of the QAOA circuit using the XY-Hamiltonian mixer and Dicke-state initialization for a two-node, three-channel system. The dynamics are confined to the tensor-product Johnson-graph subspace defined by the node-wise Hamming-weight constraints.}
    \label{fig:circuit_process}
\end{figure}

Figure~\ref{fig:w-state} shows a circuit schematic for preparing Dicke states for a two-node, three-channel example, and Fig.~\ref{fig:circuit_process} outlines the overall QAOA circuit including Dicke-state initialization and XY-mixer layers.

\subsubsection*{Johnson-graph viewpoint on the feasible subspace}

For each node $v_i$, the local channel register consists of $m$ qubits subject to the Hamming-weight constraint $\sum_{c=1}^{m} x_{i,c} = k_i$. The set of valid local configurations can be identified with the set of $k_i$-element subsets of $\{1,\dots,m\}$, which form the vertex set of the Johnson graph $J(m,k_i)$. The global set of classical configurations that satisfy all node-wise constraints is
\begin{equation}
    \mathcal{F}
    = \left\{ (S_1,\dots,S_n) \ \middle|\ S_i \subseteq \{1,\dots,m\},\ |S_i| = k_i \right\},
\end{equation}
with cardinality
\begin{equation}
    |\mathcal{F}| = \prod_{i=1}^{n} \binom{m}{k_i}.
\end{equation}
The quantum state space explored by our ansatz is the span of computational basis states labeled by $\mathcal{F}$, which one can regard as the vertex set of the tensor-product graph
\begin{equation}
    \mathcal{G}_{\mathrm{prod}} = \bigotimes_{i=1}^{n} J(m,k_i).
\end{equation}
The XY-mixer Hamiltonian induces edges in $\mathcal{G}_{\mathrm{prod}}$ by exchanging excitations within each node, and therefore implements a continuous-time quantum walk on this graph. The cost Hamiltonian $H_P$ assigns energies to vertices of $\mathcal{G}_{\mathrm{prod}}$ according to the number of interference conflicts, without leaving the feasible tensor-product Johnson-graph subspace.

\subsubsection*{Relation to prior XY-mixer constructions}

Our mixer Hamiltonian $H_{XY}$ is inspired by Hamming-weight preserving
XY mixers introduced for one-hot encodings in graph coloring and related
problems\cite{WangXYMixer}. In most of these works, the constraint takes the
form of a single global Hamming weight (for example, selecting exactly $B$ out of $n$
assets in portfolio optimization, or enforcing one color per vertex), and the
XY mixer acts on all qubits subject to that global constraint\cite{HeXYPortfolio,BrandhoferPortfolio,NiroulaXYQAOA}.

In the CBRS multi-channel allocation problem, the natural constraint structure is different. Each CBSD $v_i$ is associated with a local register of $m$ channel qubits and its own demand $k_i$, and the feasible subspace is the tensor product of node-local Dicke subspaces with heterogeneous weights $\{k_i\}$. We therefore construct $H_{XY}$ as a sum of intra-register XY couplings such that the evolution preserves the Hamming weight of each node individually rather than only the total number of excitations. From the perspective of QAOA, our ansatz can be viewed as a block-diagonal generalization of XY-mixer QAOA, tailored to a multi-register, multi-coloring setting with CBRS-specific interference structure.

\subsubsection*{Algorithmic description}

Algorithm~\ref{alg:subspace-qaoa} summarizes the proposed subspace-confined QAOA procedure for CBRS multi-channel assignment.

\begin{algorithm}[H]
\caption{Subspace-confined QAOA for CBRS multi-coloring}
\label{alg:subspace-qaoa}
\begin{algorithmic}[1]
\Require Interference graph $G=(V,E)$ with $|V|=n$, number of channels $m$, node demands $\{k_i\}_{i=1}^n$, QAOA depth $p$
\State Encode each node $v_i \in V$ as $m$ qubits $\{(i,1),\dots,(i,m)\}$ with one-hot channel encoding.
\State Construct the problem Hamiltonian $H_P$ that counts interference conflicts between adjacent nodes, without penalty terms for the demand constraints.
\For{$i = 1$ to $n$}
    \State Prepare the Generalized Dicke state $|D_{k_i}^{m}\rangle$ on the $m$-qubit register of node $v_i$ using a deterministic circuit.
\EndFor
\State Set the global initial state as $|\psi_0\rangle = \bigotimes_{i=1}^{n} |D_{k_i}^{m}\rangle_{v_i}$.
\State Define the XY-mixer Hamiltonian
\Statex \hspace{\algorithmicindent}$H_{XY} = \dfrac{1}{2}\displaystyle\sum_{i=1}^{n}\sum_{(c,c')\in E_{\mathrm{mix}}^{(i)}} 
    \bigl(\sigma_{(i,c)}^{x}\sigma_{(i,c')}^{x} + \sigma_{(i,c)}^{y}\sigma_{(i,c')}^{y}\bigr),$
\Statex where $E_{\mathrm{mix}}^{(i)}$ denotes intra-node qubit pairs.
\State Initialize variational parameters $\boldsymbol{\gamma}, \boldsymbol{\beta} \in \mathbb{R}^{p}$.
\For{$\ell = 1$ to $p$}
    \State Apply the problem unitary: $|\psi\rangle \gets e^{-i\gamma_{\ell} H_P} |\psi\rangle$.
    \State Apply the mixer unitary: $|\psi\rangle \gets e^{-i\beta_{\ell} H_{XY}} |\psi\rangle$.
\EndFor
\State Measure the final state $|\psi(\boldsymbol{\gamma},\boldsymbol{\beta})\rangle$ in the computational basis to obtain a bitstring $z$.
\State Evaluate the classical objective $C(z)$ (conflict count) and update $(\boldsymbol{\gamma},\boldsymbol{\beta})$ using a classical optimizer.
\State Repeat the quantum--classical optimization loop until a stopping criterion is met.
\Ensure A bitstring $z^{\star}$ corresponding to a multi-coloring that approximately minimizes interference while exactly satisfying $\sum_{c} x_{i,c} = k_i$ for all $v_i\in V$.
\end{algorithmic}
\end{algorithm}

\subsection*{Dual-constraint QAOA ansatz}

In some CBRS deployment scenarios, not only the per-node channel
demands $k_i$ but also the per-channel capacities $L_c$ are fixed by
regulatory, hardware, or coexistence requirements. In that case,
feasible channel-allocation matrices $X \in \{0,1\}^{n \times m}$
must satisfy two sets of cardinality constraints,
\begin{align}
    \sum_{c=1}^{m} x_{i,c} &= k_i,
    && \forall i \in \{1,\dots,n\}, \label{eq:dual_node}\\
    \sum_{i=1}^{n} x_{i,c} &= L_c,
    && \forall c \in \{1,\dots,m\}, \label{eq:dual_channel}
\end{align}
in addition to the interference constraints encoded in $H_P$. We refer
to this setting as dual-constraint multi-channel allocation,
since both node-wise demands and per-channel capacities must hold
simultaneously.

\subsubsection*{Deterministic dual-feasible initialization}

To construct an initial quantum state supported entirely on the
intersection of the node-wise and channel-wise cardinality
subspaces, we first build a canonical binary matrix
$X^{(0)} \in \{0,1\}^{n \times m}$ that satisfies
Eqs.~\eqref{eq:dual_node}--\eqref{eq:dual_channel}. In our implementation,
we employ a simple greedy fill: starting from the zero matrix, we traverse
nodes and channels and place ones whenever both the remaining node demand $k_i$ and
the remaining channel capacity $L_c$ are positive. This procedure
terminates with
\(
\sum_c X^{(0)}_{i,c} = k_i
\)
for all $i$ and
\(
\sum_i X^{(0)}_{i,c} = L_c
\)
for all $c$, provided that
$\sum_i k_i = \sum_c L_c$.

On the quantum register, we prepare the corresponding computational
basis state
\[
    \ket{X^{(0)}} = \bigotimes_{i=1}^n \bigotimes_{c=1}^m
    \ket{x^{(0)}_{i,c}},
\]
by applying single-qubit $X$ gates to all qubits $(i,c)$ such that
$X^{(0)}_{i,c} = 1$. In contrast to the Dicke-state initialization
used in our main ansatz, this dual-constraint variant starts from a
single strictly feasible configuration that satisfies both
Eqs.~\eqref{eq:dual_node} and \eqref{eq:dual_channel}.

\subsubsection*{Plaquette-based dual mixer}

To preserve both sets of constraints during the QAOA evolution, we
introduce a local four-qubit plaquette mixer that acts on pairs of
nodes and pairs of channels. For two nodes $i,j$ and two channels
$c \neq c'$, we consider the four qubits
\[
    (i,c),\ (i,c'),\ (j,c),\ (j,c'),
\]
and label the corresponding computational basis states as
$\ket{b_1 b_2 b_3 b_4}$. Among the 16 possible configurations, the two
patterns
\[
    \ket{1001} \quad\text{and}\quad \ket{0110}
\]
represent an exchange of channel allocations $c$ and $c'$ between nodes $i$
and $j$ while preserving both row sums and column sums in the $2\times 2$
submatrix spanned by $(i,j) \times (c,c')$.

We define a local plaquette Hamiltonian
\begin{equation}
    H_{\square}^{(i,j;c,c')}
    =
    \ketbra{1001}{0110}
    +
    \ketbra{0110}{1001},
\end{equation}
which acts as a Pauli-$X$ operator in the two-dimensional subspace
spanned by $\{\ket{1001},\ket{0110}\}$ and as the identity on all
other basis states. The corresponding unitary
\begin{equation}
    U_{\square}(\beta)
    = \exp\bigl(- i \beta H_{\square}^{(i,j;c,c')}\bigr)
\end{equation}
implements a simple rotation between $\ket{1001}$ and $\ket{0110}$ while leaving all other configurations unchanged. By construction,
$U_{\square}(\beta)$ preserves the row sums
$\sum_c x_{i,c}$, $\sum_c x_{j,c}$ and the column sums
$\sum_i x_{i,c}$, $\sum_i x_{i,c'}$, and therefore maps any globally
feasible configuration to another globally feasible configuration.

Aggregating these local couplings over all pairs of nodes and channels
yields the dual-constraint mixer Hamiltonian
\begin{equation}
    H_{\mathrm{mix}}^{\mathrm{dual}}
    =
    \sum_{1 \le i < j \le n}
    \sum_{1 \le c < c' \le m}
    H_{\square}^{(i,j;c,c')},
\end{equation}
and the corresponding mixer unitary
$\exp(- i \beta H_{\mathrm{mix}}^{\mathrm{dual}})$ defines a quantum
walk on the connected component of the feasible
subspace that contains $\ket{X^{(0)}}$.

\subsubsection*{Dual-constraint QAOA layer}

A depth-$p$ dual-constraint QAOA state is given by
\begin{equation}
    \ket{\psi_p^{\mathrm{dual}}(\boldsymbol{\gamma},\boldsymbol{\beta})}
    =
    \prod_{\ell=1}^{p}
    \exp\bigl(-i \beta_{\ell} H_{\mathrm{mix}}^{\mathrm{dual}}\bigr)
    \exp\bigl(-i \gamma_{\ell} H_{P}\bigr)
    \ket{X^{(0)}},
\end{equation}
where $H_P$ is the same conflict-counting problem Hamiltonian used in
the Dicke+XY ansatz. Each layer preserves both the node-wise demands
in Eq.~\eqref{eq:dual_node} and the channel-wise capacities in Eq.~\eqref{eq:dual_channel}, so the optimization is restricted
to the intersection of the corresponding cardinality subspaces. In our
numerical study, we implement the plaquette unitaries
$U_{\square}(\beta)$ as explicit four-qubit unitaries acting on the
computational basis, and we focus on shallow circuits with $p=1$ to
highlight the qualitative behavior of the dual-constraint mixer on
small CBRS instances.

\subsection*{Classical baselines}

To contextualize the performance of our subspace-confined QAOA ansatz, we compare
it against two classical baselines for graph multi-coloring.

\paragraph{Exact ILP formulation.}
For moderate-size instances (up to $n=8$ nodes and $m=3$ channels),
we formulate the CBRS multi-channel allocation problem as a 0--1 integer linear
program (ILP). The binary decision variables $x_{i,c} \in \{0,1\}$ indicate whether channel $c$ is assigned to node $v_i$. The ILP enforces the demand and interference constraints
\begin{align}
    \sum_{c=1}^{m} x_{i,c} &= k_i,
    && \forall v_i \in V, \label{eq:ilp_demand}\\
    x_{i,c} + x_{j,c} &\le 1,
    && \forall (v_i,v_j) \in E,\ \forall c \in \{1,\dots,m\}, \label{eq:ilp_interference}
\end{align}
and minimizes the total number of interference conflicts,
\begin{equation}
    \min_{x \in \{0,1\}^{n m}} 
    C(x)
    =
    \sum_{(v_i,v_j) \in E}
    \sum_{c=1}^{m} x_{i,c} x_{j,c}.
\end{equation}
Equivalently, one can solve a feasibility problem with the objective set to zero whenever Eqs.~\eqref{eq:ilp_demand}--\eqref{eq:ilp_interference} are satisfied. We use a standard off-the-shelf solver (such as Gurobi or CBC) as a reference for the globally optimal classical solution on small CBRS interference graphs.

\paragraph{Greedy multi-coloring heuristic.}
For larger or repeated instances, we also consider a simple greedy multi-coloring heuristic inspired by DSATUR-type graph coloring algorithms. The heuristic maintains, for each node $v_i$, the set of channels already assigned to its neighbors and iteratively selects node--channel pairs according to a priority rule. At each step, it chooses the node with the largest residual demand (breaking ties by degree or at random), selects a channel that does not violate current interference constraints and optimizes a local score (for example, least-used across neighbors), assigns that channel, and updates the residual demands and neighbor channel sets. If no conflict-free channel is available, the algorithm either leaves part of the demand unsatisfied or accepts a conflict, depending on the evaluation scenario. This heuristic is computationally inexpensive and provides a useful classical baseline, but it does not guarantee optimality and can be trapped in high-conflict configurations, especially for dense graphs with heterogeneous multi-channel demands.

\subsection*{Simulation setup}

All simulations were implemented in Python using Qiskit. For the 8-node experiment, the interference graph $G$ was generated as an 8-node ring with additional cross edges chosen to mimic realistic interference patterns among CBSDs. The number of available channels was set to $m=3$, and the per-node demands were chosen as $k = [2, 1, 2, 1, 1, 2, 1, 1]$, corresponding to heterogeneous PAL requirements. This configuration leads to $N_{\mathrm{qubits}} = 24$.

For ideal (noise-free) experiments, we used Qiskit’s \texttt{AerSimulator} in statevector mode and generated measurement outcomes with a finite number of shots (typically 1{,}024) to emulate NISQ conditions. The classical optimization of QAOA parameters $(\boldsymbol{\gamma},\boldsymbol{\beta})$ was performed using the COBYLA optimizer with a maximum number of iterations between 30 and 80, depending on the instance. The same set of initial parameters was used for both the standard and proposed ansätze in each experiment to ensure a fair comparison.

For the noise-robustness study on a 5-node (15-qubit) instance, we employed a depolarizing noise model applied to both single- and two-qubit gates, with error rates $p_{\mathrm{err}}$ ranging from 0 to 5\%. For each noise level, we executed the QAOA circuits multiple times and estimated the average deviation from the Hamming-weight constraints based on histograms accumulated over several thousand measurement shots, as described in the Results section. The reported deviation values in Fig.~\ref{fig:performance}(b) correspond to averages over these repeated runs.

\bibliography{sn-bibliography_QAOA}

\section*{Acknowledgements}

\section*{Author contributions}

GSM conceived the proposed method. GSM and YJS performed the simulations, derived the analytical equations, and prepared the figures. GSM, YJS and JH jointly developed the core ideas, and JH supervised the project.

\section*{Data availability}
    
The datasets generated during and/or analysed during the
current study are available from the corresponding author on reasonable request.

\section*{Competing interests}
The authors declare no competing interests.

\section*{Funding}
The authors did not receive support from any organization for the submitted work. No funding was received to assist with the preparation of this manuscript.

\end{document}